\documentclass[preprint,aps,superscriptaddress,nofootinbib]{revtex4}

\usepackage{amsmath}    
\usepackage{graphicx}  
\usepackage{natbib}
\usepackage{cases}
\usepackage{color}
\usepackage{slashed}

\begin{document}

\title{ Classical and Quantum Szilard Engine under Generalized Uncertainty Principle Effect }
\author{Chih-Wei Chen}\thanks{
E-mail:rty17282004@gmail.com }
\affiliation{Department of Physics and Center for High Energy Physics, Chung Yuan Christian University, Chung Li District, Taoyuan, Taiwan}

\author{Wen-Yu Wen}\thanks{%
E-mail: steve.wen@gmail.com}
\affiliation{Department of Physics and Center for High Energy Physics, Chung Yuan Christian University, Chung Li District, Taoyuan, Taiwan}
\affiliation{Leung Center for Cosmology and Particle Astrophysics\\
National Taiwan University, Taipei 106, Taiwan}

\begin{abstract}
We studied the Szilard engine under the effect of generalized uncertainty principle (GUP).  In the classical Szilard engine, the work done by the engine is reduced by the GUP effect via a modified ideal gas law.  In the quantum Szilard engine, the correction comes from the shifted eigen energy due to the nonlinear momentum dependence.  We studied its effect on both bosonic and fermionic molecules.
\end{abstract}

%\date{\today}
\maketitle

\section{Introduction}

The thought experiment of Maxwell's demon \cite{Leff03,Maruyama09} was to challenge the second law of the thermodynamics, which states that the entropy can never decrease as time evolves.  The Szilard engine \cite{Szilard29}, as a working example, provides us control handles to realize this daring dream in a quantitative and precise way.  The simplest model of classical Szilard Engine (cSZE) consists of a frictionless movable wall, empty box in which a single gas molecule resides, and a weight to extract the work.  The Szilard Engine operates in a mostly isolated environment with four main phases: first inserting the wall to divide the box into two equal parts, then the particle's location is measured to determine which side to hang the weight.  The isothermal expansion of gas lifts the weight while in contact with a thermal bath.  At last removing the wall to complete a cycle of the Szilard Engine.  The engine is running based on the information from position measurement of the gas particle.

If we carefully insert and remove the wall in steady speed and exclude the work done by the demon in order to measure and position, the work done by the engine is simply due to the expanding phase, that is
\begin{equation}\label{W=ln2}
W_c=k_B T_{bath}\ln{2}, %1
\end{equation}
where $T_{bath}$ is the temperature of thermal bath.   In contrast, the quantum Szilard engine (qSZE), where the box is regarded as the infinite potential well, does work in all three phases: insertion, expansion and removing\cite{Kim:2011zzc}.  For a single particle, there appears no difference on the total work from the classical result.  The quantum effect comes into play for more than one particle.  In particular, the total work done by a two-particle qSZE can be expressed as
\begin{equation}
W_q = -2k_B T_{bath} f_0 \ln f_0,
\end{equation}
where at low temperature, $f_0$ approaches $1/3$ for two identical bosons, but $0$ for fermions.  Both approach classical work ($f_0=1/4$) as $T_{bath} \to \infty$.

On the other hand, classical thermodynamics is about to break down for a runaway temperature as the Schwarzschild black hole evaporates via the Hawking radiation, that is 
\begin{equation}
T_{BH} = \frac{\hbar c^3}{8\pi GM k},
\end{equation}
and a UV complete theory of quantum gravity is expected to emerge at the Planck scale or string scale \cite{Konishi:1989wk}.  However, a convincing theory with supporting evidences from observation or experiments is unavailable at the moment or at least difficult.  An economic and effective way to generate a new scale of the order of Planck length is via the generalized uncertainty principle (GUP) satisfying the commutative relation \cite{Maggiore:1993rv}:
\begin{equation}
[\hat{x},\hat{p}]=i\hbar(1+\alpha \hat{p}^{2}) %6
\label{eqn_commutative}
\end{equation}
Application of this modification to the black hole thermodynamics implies a possible existence of remnants, which in terms set up a scale for maximum temperature in our universe \cite{Adler:2001vs,Scardigli:2010gm}.   It is interesting to investigate its application to other thermodynamic systems.  In particular, it is unclear if the Szilard engine does work differently under the effect of GUP.  Will the modification of GUP cause any form of violation of the second law?  Does the GUP have same or different effect on bosons and fermions?  In this letter we will try to find answers for those questions.
 
This paper is organized as follows: we review the GUP-motivated ideal gas law in two spatial dimensions and derive the correction to the work done by a cSZE in the section II.  We found the engine does less work due to the GUP effect.  In the section III, we discuss the work done by the qSZE for identical bosons and fermions.  We discovered that their statistical behaviors will be altered and quantum aspects can already been spotted at slightly higher temperature.

\section{GUP-modified ideal gas law and classical Szilard engine}
For simplicity, we will consider the cSZE made of a two-dimensional box filled with ideal gas.  The commutative relation (\ref{eqn_commutative}) implies the following uncertainty principle:

\begin{equation}\label{GUP}
\Delta x \Delta p \geq \frac{\hbar}{2}[1+\alpha (\Delta p)^{2}] %7
\end{equation}

or equivalently a suppression at high momentum in the phase space

\begin{equation}\label{density of state}
\frac{2dxpdp}{h^{2}(1+\alpha p^{2})}%11
\end{equation}

Incorporating the GUP effect into the kinetic theory of non-relativistic gas respecting the Maxwell-Boltzmann statistics, the differential of number density $n(p)$ reads,

\begin{equation}\label{density_GUP}
\frac{d n(p)}{dp} \sim \frac{2\pi pexp(\frac{-p^{2}}{2mk_{B}T})}{h^{2}(1+\alpha p^{2})^{2}},%12
\end{equation} 
where  $m$ is the mass of gas molecule.  This leads to the modified ideal gas law up to ${\cal O}(\alpha)$ correction:
 
\begin{equation}
PV=(1-\alpha^\prime k_{B}T)Nk_{B}T + {\cal O}(\alpha^{2}), %13
\label{eqn_ideal_gas_law}
\end{equation}
where $\alpha^\prime=12\alpha m$ in the nature unit $h=1$.  We leave the derivation in the appendix.

\begin{figure}
\centering
\includegraphics[scale=0.8]{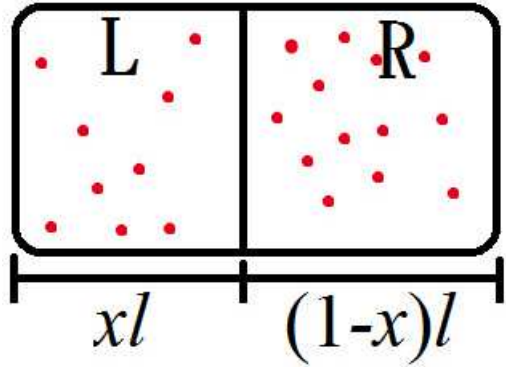}
\caption{2D CSE model, the box is filled with ideal gas and separated by movable wall.  The sizes of left and right compartment are $x\l$ and $(1-x)\l$ respectively.  We consider the initial state with $x=1/2$.}
\label{2D-CSE model}
\end{figure}

Now we consider the configuration as shown in the Figure \ref{2D-CSE model}, where there are $n$ gas molecules at the left compartment and $N-n$ particles at the right.  During the isothermal expansion phase, the wall starts at the middle of box and moves toward the right or left upon the distribution $n/N<1/2$ or $n/N >1/2$.  The expectation value of work done by the cSZE is calculated in the appendix and has the following form \cite{Yavilberg} :

\begin{equation}\label{eqn_cSZE work}
<W>=(1-\alpha^\prime k_{B}T)\frac{k_{B}T}{2^{N}}\sum_{n=0}^{N}\binom{N}{n}\ln{[2^{N}(\frac{n}{N})^{n}(1-\frac{n}{N})^{N-n}]}.
\end{equation}

\begin{figure}
\centering
\includegraphics[scale=0.8]{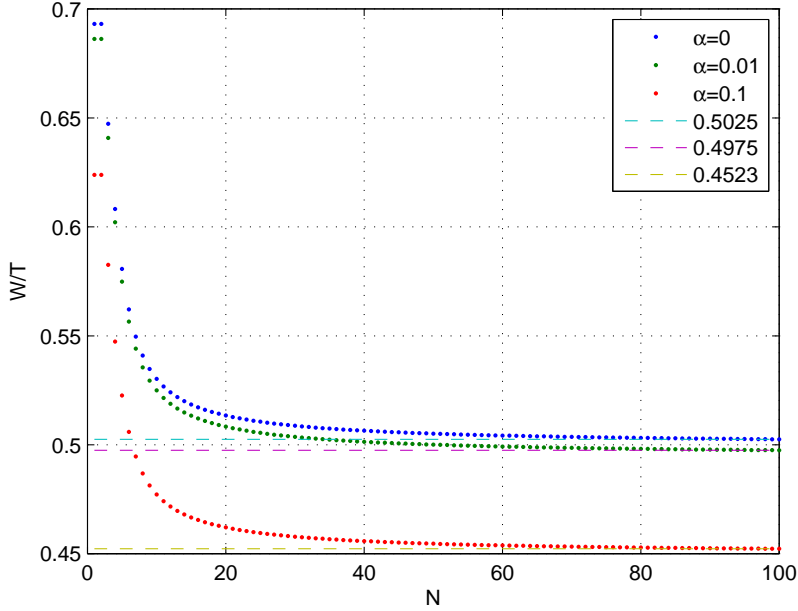}
\caption{Work done by cSZE versus particle number $N$ for different $\alpha$.  (a) When $\alpha=0$, $W/T=0.6931$ at $N=1,2$ and approaches $\ln2=0.5025$ at large $N$, (b) When $\alpha=0.1$ $W/T=0.6862$ at $N=1,2$ and approaches  $W=0.4523$ at large $N$, (c) When $\alpha=0.01$, $W/T=0.6862$ at $N=1,2$ and approaches $0.4975$ at large $N$.}
\label{cSZE}
\end{figure}

We plot it for various $\alpha$ in the Figure \ref{cSZE}.  It shows that the GUP effect reduces the work done by cSZE.  With same amount of volume change, the average pressure decreases due to suppression of number density at large momentum.

Now we are ready to investigate the GUP effect on the Second Law.  Without loss of generality, we can consider a single gas molecule in the 2D box.  The work done by the cSZE during the isothermal expansion at temperature $T_{bath}$ is given by 

\begin{equation}
W_c^{\prime}=\int \frac{dQ}{T_{bath}}=\int_{\frac{L}{2}}^{L-\sqrt{\alpha}\hbar} \frac{PdV}{T_{bath}}
\end{equation}
where we have considered the minimal length scale $\sqrt{\alpha}\hbar$ under which the compression is impossible.  Using (\ref{eqn_ideal_gas_law}), one achieves
\begin{equation}
W_c^{\prime} \simeq k_BT_{bath}( 1-12m\alpha^\prime k_B T_{bath}) \ln[2(1-\frac{\sqrt{\alpha}\hbar}{L})] + {\cal O}(\alpha^2).
\end{equation}

We have the following remarks: at first, the above work formulas reduces to (\ref{W=ln2}) at the commutative limit where $\alpha \to 0$.  Secondly, it is obvious that $W_c^{\prime} < W_c=k_BT_{bath}\ln 2$.  If it would take the same amount of entropy for the {\sl demon} to erase its old memory\cite{Leff03}, say $W_c$, then the net change of entropy would be $\Delta S = -W_c^{\prime} + W_c >0$, where the Second Law is satisfied.

\section{quantum Szilard Engine}

\begin{figure}
\includegraphics[scale=0.8]{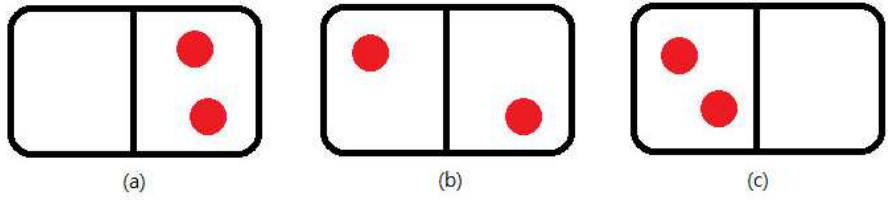} 
\caption{There are three configurations for the case of two particles.  We define the function $f_i$ as the probability to find $i$ particle(s) in left compartment.   For example, the probability is (a) $f_{0}$ if there is no particle in left compartment, (b) $f_{1}$ if there is one particle in left compartment, (c) $f_{2}$ if there are two particles in left compartment.}
\label{two identical particles}
\end{figure}

For the qSZE, we treat the particles in the box as the problem of infinite potential well of width $X$.  The particle in the box is described by a wave function  satisfying the Schr{\"o}dinger equation with GUP correction:

\begin{equation}\label{TISE 2}
-\frac{\hbar^{2}}{2m}\frac{d^{2}\psi}{dx^{2}}+\frac{\alpha\hbar^{4}}{m}\frac{d^{4}\psi}{dx^{4}}+V(x)\psi  = E\psi
\end{equation}

The wave fuction can be solved and energy eiigenvalues are\cite{Blado:2014}

\begin{equation}\label{energy modication}
E_{n}=\frac{n^{2}\pi^{2}\hbar^{2}}{2mX^{2}}+\alpha\frac{n^{4}\pi^{4}\hbar}{mX^{4}}
\end{equation}

While the dividing wall moves from $X_1$ to $X_2$, the work done by qSZE is given by\cite{Kim:2011zzc} 

\begin{equation}
W=k_{B}T_{bath}\sum_{n}\int_{X_{1}}^{{X_{2}}}\frac{\partial \ln{Z}}{\partial E_{n}}\frac{\partial E_{n}}{\partial X}dX=k_BT_{bath}\ln\frac{Z(X_2)}{Z(X_1)},
\end{equation}

where the partition function is

\begin{equation}\label{modication partition function}
Z(X,\beta)=\sum_{n}exp[-\beta (\frac{n^{2}\pi^{2}\hbar^{2}}{2mX^{2}}+\alpha\frac{n^{4}\pi^{4}\hbar}{mX^{4}})]
\end{equation}

We will focus on the least nontrivial case of two identical particles and their partition is shown in the figure  \ref{two identical particles}.  In particular, we compare the function $f_0$ for bosons and fermions.  At high temperature limit, the quantum effect is overwhelmed by thermal fluctuation and $f_0$ approaches its classical value $\frac{1}{4}$.  At zero temperature limit, $f_0 \to \frac{1}{3}$ for bosons due to the indistinguishability of two particles.  On the other hand, $f_0 \to 0$ for fermions due to the Pauli exclusive principle.  At the finite temperature, $f_0$ can be computed using  (\ref{fn}).   It is convenient to define a new variable $d=Z(X,\beta)^2/Z(X,2\beta)$, such that $f_0$ has the expression\cite{Kim:2011zzc}

 \begin{figure}
\centering
\includegraphics[scale=0.8]{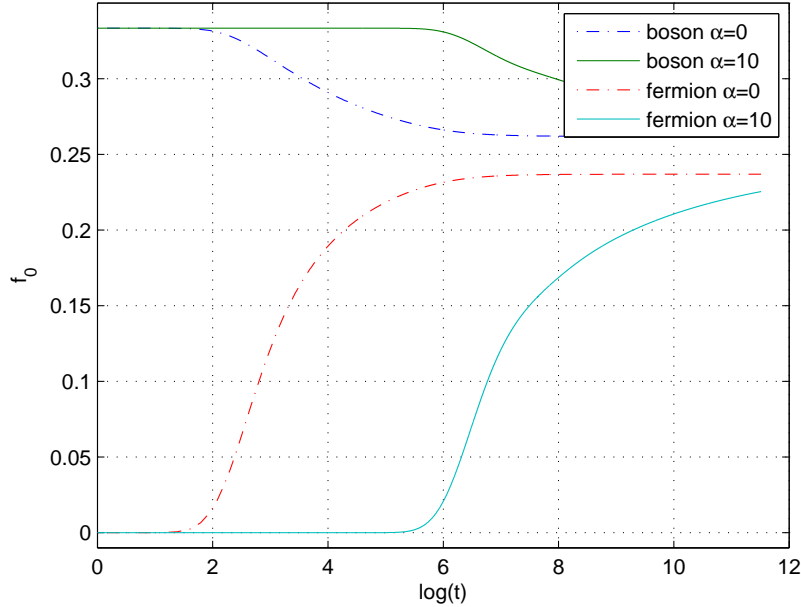} 
\caption{$f_0$ vs. temperature. The curves of dashed purple ($\alpha=0$) and solid green ($\alpha=10$) are for bosons; curves of dashed red ($\alpha=0$) and solid blue ($\alpha=10$) are for fermions.  In reality, $\alpha$ should be very small but here we exaggerate its value in order to spot the effect.}
\label{qSZE curve}
\end{figure}

\begin{numcases}{}
\frac{d+1}{4d+2},\quad \text{bosons}\\
\frac{d-1}{4d-2},\quad \text{fermions}
\end{numcases}
such that $d\to 1$ at zero temperature limit and $d\to \infty$ at high temperature.  We plot $f_0$ against temperature for bosons and fermions in the figure \ref{qSZE curve}.   We discover that as temperature decreases, the transition from classical into quantum regime appears earlier under the GUP effect.  

\begin{acknowledgments}
We are grateful to Li-Yi Hsu for his helpful discussion.  This work is supported in parts by the Chung Yuan Christian University, the Taiwan's Ministry of Science and Technology (grant No. 102-2112-M-033-003-MY4) and the National Center for Theoretical Science. 
\end{acknowledgments}

\appendix

\section{Derivation of ideal gas law}

We will derive the modified ideal gas law in equation (\ref{eqn_ideal_gas_law}).  We apply the Taylor's expansion to equation (\ref{density_GUP}) up to order $\cal O(\alpha)$:

\begin{equation}\label{density_Taylor}
\frac{dn(p)}{dp}=2\pi C pe^\frac{-p^2}{2mk_B T}(1-2\alpha p^2),%(2)
\end{equation}
where $C$ is the normalized coefficient to be determined.  This can be integrated into

\begin{equation}
n(p)=C[2\pi k_B T(1+4\alpha mk_B T)],%(5)
\end{equation}
or equivalently
\begin{equation}\label{k_value}
C=\frac{n}{2\pi k_B T(1+4\alpha mk_B T)}%(6)
\end{equation}

Then the gas pressure can be obtained by inserting (\ref{density_Taylor}) and (\ref{k_value}) into
\begin{equation}
P=\int_{0}^{\infty}\frac{p^2}{2m}(\frac{dn(p)}{dp})dp%(7)
\end{equation}
Recalling the density $n=N/V$ for particle number $N$ and occupied volume $V$, one derives the modified gas law :

\begin{equation}
PV=Nk_B T\frac{1-8\alpha mk_B T}{1+4\alpha k_B T} \simeq (1-12\alpha mk_B T)Nk_B T + {\cal O}(\alpha^2) %(16)
\end{equation}
which can be inverted into the modified temperature
\begin{equation}
T\simeq \frac{PV}{Nk_B}(1+\frac{12\alpha PV}{Nk_B}) + {\cal O}(\alpha^2).
\end{equation}

\section{Derivation of work formula for classical Szilard engine}

We will derive the expectation value of work for classical Szilard engine (\ref{eqn_cSZE work}).   We denote the partition function of $n$ particles in the left compartment of length $X$ and $N-n$ particles in the right compartment of length $L-X$ as 
\begin{equation}
Z_{n,N-n}(X) =Z_n(X) Z_{N-n} (L-X),
\end{equation}
where $Z_n(X)$ is the partition function for $n$ particles in a box of width $X$.  The above product form implies no interaction across the wall, which divides the box.  If the partition function takes the following form 
\begin{equation}
Z_n(X) = \frac{X^n}{\lambda_T n!}
\end{equation}
for some constant $\lambda_T$,  then the probability of finding $n$ particles out of $N$ in one of the compartments of length $\l_n$ has following expression: 
\begin{equation}\label{fn}
f_n = \frac{Z_{n,N-n}(\l_n)}{\sum_{n^\prime=0}^N Z_{n^\prime,N-n^\prime} (\l_n)} .
\end{equation}
In particular, if one fixes $\l_n=L/2$, then one has 
\begin{equation}\label{pn}
p_n = f_n|_{\l_n=\frac{L}{2}} =\frac{Z_{n,N-n}(L/2)}{\sum_{n^\prime=0}^N Z_{n^\prime,N-n^\prime} (L/2)} = \frac{1}{2^N}\binom{N}{n},
\end{equation}
which agrees with what is expected according to the theory of permutation.  The expectation work performed in each cycle is given by 
\begin{equation}
<W> = -(1-\alpha^\prime k_B T_{bath})k_BT_{bath} \sum_{n=0}^{N}p_n \ln(\frac{p_n}{f_n}),
\end{equation}
where the wall is initially inserted in the middle and moves toward the equilibrium position $\l_n = \frac{n}{N}\l$.  After inserting (\ref{fn}) and (\ref{pn}), one can obtain (\ref{eqn_cSZE work}).

\end{document}